\documentclass[english]{article}
\usepackage[T1]{fontenc}
\usepackage[latin9]{inputenc}
\usepackage{geometry}
\usepackage{babel}
\usepackage{amsmath}
\usepackage{amssymb}
\usepackage[unicode=true,pdfusetitle,
 bookmarks=true,bookmarksnumbered=false,bookmarksopen=false,
 breaklinks=false,pdfborder={0 0 1},backref=section,colorlinks=false]
 {hyperref}
\usepackage{breakurl}

\makeatletter

\usepackage{fullpage}\usepackage[justification=centering]{caption}\usepackage[all]{hypcap}\usepackage{microtype}

\makeatother

\begin{document}

\title{Magnetic $g$-Factors with a Surface Delta Interaction}

\author{Larry Zamick$^{1}$, Brian Kleszyk$^{1}$, Yitzhak Sharon$^{1}$,
Shadow Robinson$^{2}$ \\
 \textit{$^{1}$Department of Physics and Astronomy, Rutgers University,
Piscataway, New Jersey 08854, USA} \\
 \textit{$^{2}$Department of Physics, Millsaps College,Jackson,Mississippi,
39210, USA} }

\date{\today}
\maketitle
\begin{abstract}
Using an attractive surface delta interaction we obtain wave functions
for 2 protons (or proton holes) in the $f_{5/2}$ and $p_{3/2}$ shell.
We take the single particle energies to be degenerate. We obtain the
remarkable result that the magnetic $g$-factors of the lowest $2^{+}$
state is equal to 1 and the same is true for the lowest $4^{+}$ state.
Only the orbital part of the $g$-factors contribute $-$ the spin
part cancels out. Then shell model calculations are performed for
these same g factors in a larger space and with realistic effective interactions. 
\end{abstract}

\section{The Surface Delta Interaction}

The surface delta interaction (SDI) of Green and Moszkowski \cite{Green}
and Arvieu and Moszkowski \cite{Arvieu} has proven to be a very useful
schematic interaction. It can be used to find the hidden simplicity
in complex calculations. This interaction has been extensively discussed
in Talmi's book \cite{Talmi}.

The matrix element of the SDI interaction can be written as follows:
\begin{equation}
<[j_{1}j_{2}]SDI[j_{3}j_{4}]>=C_{0}f(j_{1},j_{2})f(j_{3},j_{4})
\end{equation}
where we have 
\begin{equation}
f(j_{1},j_{2})=(-1)^{j_{2}+\frac{1}{2}}\sqrt{\frac{2(2j_{1}+1)(2j_{2}+1)}{(2J+1)(1+\delta_{j_{1}j_{2}})}}\times\Big<j_{1}j_{2}\text{\space}\frac{1}{2}\left(\frac{\text{-}1}{2}\right)\Big|J0\Big>
\end{equation}
 Note that the expression is separable and so, as indicated by Talmi
\cite{Talmi} it is easy to obtain the lowest state wave functions
(see Eq. 12.49). In our notation we use 
\begin{equation}
\psi^{J}=\sum f(j_{1},j_{2})[j_{1}j_{2}]^{J}
\end{equation}

\section{Wave Functions and $g$-Factors}

We consider 2 proton particles or 2 proton holes in either the $f_{5/2}$
or $p_{3/2}$ orbitals. As an orientation the 2 hole system could
be $^{86}$Kr. We take the single particle energies to be degenerate.
We choose $C_{0}$ to be negative. We find that the wave function
of the $J=4^{+}$ state is 
\begin{equation}
a[f_{5/2}f_{5/2}]^{4}+b[f_{5/2}p_{3/2}]^{4}
\end{equation}
 and similarly the wave function of the $2^{+}$ state is 
\begin{equation}
c[f_{5/2}f_{5/2}]^{2}+d[f_{5/2}p_{3/2}]^{2}+e[p_{3/2}p_{3/2}]^{2}
\end{equation}
 The following values of the coefficients are obtained by matrix diagonalization:
\begin{align*}
 & J=4^{+}\rightarrow a=-0.447189,\text{\space}b=0.894444\\
 & J=2^{+}\rightarrow c=0.692819,\text{\space}d=-0.48992,\text{\space}e=0.5291337
\end{align*}
 Using the bare $g$-factors for the protons (g$_{l}$=1, g$_{s}$
=5.586) the values of g(J) for the 2 hole system are: 
\begin{align*}
g(4^{+}) & =0.3448a^{2}+1.1638b^{2}\\
g(2^{+}) & =0.3448c^{2}+0.5268d^{2}+2.5287e^{2}
\end{align*}
 With the above values we find the interesting result that g(2$^{+}$)
is equal to g(4$^{+}$) and both are equal to one or more properly
stated both are equal to $g_{l}$.

To gain further insight we did calculations with $g_{l}=0$ but keeping
g$_{s}$ unchanged. The expressions are 
\begin{align*}
g(4^{+}) & =-0.798a^{2}+0.1995b^{2}\\
g(2^{+}) & =-0.798c^{2}-0.57633333d^{2}+1.862e^{2}
\end{align*}
 We find that both $g(4^{+})$ and $g(2^{+})$ vanish when we set
$g_{l}=0$ i.e. the spin contributions cancel out. If we set $g_{s}=0$
i.e. we include only the orbital part and the expressions become 
\begin{align*}
g(4^{+}) & =1.14285714286a^{2}+0.9642857144b^{2}\\
g(2^{+}) & =1.14285714286c^{2}+1.103188889192d^{2}+0.66666667e^{2}
\end{align*}
 and one sees that again $g(4^{+})=g(2^{+})=1$. This behavior is
connected with a combination of pseudo-spin symmetry developed by
Hecht et al. \cite{Hecht} and Arima et al. \cite{Arima}. Insights
into the origin of pseudospin symmetry via the Dirac equation has
been giving by Ginocchio and Leviatan \cite{Ginocchio}. Also relevant
is the quasi-spin formulation of Kerman \cite{Kerman} and Talmi's
generalized seniority scheme \cite{Talmi}. From the pseudo-spin formulation
one can map the orbits $f_{5/2}$ and $p_{3/2}$ into $d_{5/2}$ and
$d_{3/2}$. We further note that the surface delta interaction is
a quasi-spin conserving interaction. In the $d$ shell the expressions
for the $g$-factors are a bit more complicated because the magnetic
moment operator can connect $d_{5/2}$ and $d_{3/2}$. The expressions
are: 
\begin{align*}
g(4^{+}) & =1.9172a^{2}+1.2293b^{2}+0.917199968ab\\
g(2^{+}) & =1.9172c^{2}+1.764333333d^{2}+0.0828e^{2}+2.5942332913cd-1.98137832564de
\end{align*}

One can verify that with the surface delta interaction $g(4^{+})=g(2^{+})=1$.
A simpler argument in the $d$ shell is that with a spin independent
interaction (i.e. Wigner interaction) LS coupling holds. Thus the
$J=2^{+}$ state has $L=2$, $S=0$ whilst the $J=4^{+}$ state has
$L=4$, $S=0$. Hence only the orbital angular momentum contributes
to the $g$-factors. However in the $f_{5/2}$-$p_{3/2}$ space it
is far less obvious. We briefly comment on single particle energies.
In the theoretical work of Hjorth-Jensen et al. \cite{Hjorth-Jensen}
the $f_{5/2}$-$p_{3/2}$ single particle splitting is 4.4 MeV. However
this is in anticipation of a large scale shell model calculation with
many valence nucleons relative to $^{40}$Ca. If we look more locally
we note that for $^{85}$Br (one proton less that $^{86}$Kr) the
ground state has $J=\frac{3}{2}^{-}$ with the $J=\frac{5}{2}^{-}$
state at an excitation energy of 0.345 MeV. (This is in stark contrast
to the naive using of the single particle energy splitting in \cite{Hjorth-Jensen}
which would lead to a $J=\frac{5}{2}^{-}$ ground state with the $J=\frac{3}{2}^{-}$
state at 4.4 MeV.) So taking the single particle energies degenerate
is not so bad.

It should be pointed out that the experimental $g$-factor of the
$2^{+}$ state is very close to 1. In the work of T.J. Mertzimekis
et al. \cite{Mertzimekis} they quote the value 1.12 (14). Recent
measurements by Kumbartzki et el. {[}11{]} for the 2$^{+}$ gives
a value consistent with this 1.10(5). They have a new result for th
4$^{+}$ 1.07(15). 

Amusingly, both g(2$^{+}$) and g(4$^{+}$) are close to unity, as
is given by the surface delta interaction in the small space. This
by itself does not mean that the wave functions of this interaction
are literally correct but this schematic model gives us ideas of examining
the orbital ans spin contents of magnetic g factors.

Most $g$-factors of excited states of even-even nuclei are close
to $\frac{Z}{A}$ so that a $g$-factor of 1 is unusual in any circumstance.
Things that conspire to yield this result are that we have a closed
shell of neutrons $N=50$ and the closeness of the $f_{5/2}$ and
$p_{3/2}$ single particle energies. We however do not have any explicit
empirical evidence that the $g$-factor is dominantly orbital.

Although in this simple model the expectation value of $g$$_{s}$
is calculated to be zero in $^{86}$Kr this does not mean that we
can assign a quantum number $S=0$ for it. Expressed in LS coupling
in terms of basis states {[}(l$_{1}$ l$_{2}$)$^L$ (1/2 1/2)$^S${]}$^{J}$the
wave function ,is fairly complicated e.g. for $J=4^{+}$. 
\begin{equation}
\begin{array}{llll}
\Psi= & 0.07376[(33)^{3}1]^{4} & -0.21187[(33)^{4}0]^{4} & -0.38684[(33)^{5}1]^{4}\\
 & -0.16903[(31)^{3}1]^{4} & +0.58717[(31)^{4}0]^{4} & +0.6547[(31)^{4}1]^{4}
\end{array}
\end{equation}

We thus have admixtures of L=3, 4 and 5 and S=0 and 1 in this wave
function. Note that the basis state {[}(33)$^4$($\frac{1}{2}\frac{1}{2}$)$^1${]}$^{4}$ is
symmetric and therefor must be excluded.

It should also be pointed out that the probability of the $(33)$
configuration is $0.2$ and the probability of the $(31)$ is $0.8$.
The expectation value $S$ then for the first three terms ``$(33)$''
is -0.1428 and for the last three terms ``$(31)$'' is 0.1428 giving
a net value of 0. But clearly $L$ and $S$ are not good quantum numbers.
It is not a priori obvious that the above wave function would give
such a simple result for the g factors and its purely orbital character.

We next look at the experimental data. The observed energies of the
first $2^{+}$ and $4^{+}$ states in $^{86}$Kr are 1.565 MeV and
2.350 MeV respectively. However with the SDI they would be degenerate.
With $C_{0}=-0.5$ MeV they would be at an excitation energy of 1.786
MeV. We now introduces a single particle energy splitting splitting
\begin{equation}
\epsilon^{-1}(f_{5/2})-\epsilon^{-1}(p_{3/2})=0.4\text{MeV}
\end{equation}
 from the $^{87}$Rb spectrum (single hole). We are interested in
seeing how this affects the $g$-factors. We find 
\begin{align*}
 & J=4^{+}\rightarrow a=-0.297292,\text{\space}b=0.954787\\
 & J=2^{+}\rightarrow c=0.368227,\text{\space}d=-0.390574,\text{\space}e=0.84378
\end{align*}
 which gives us $E(2^{+})=1.684$ MeV and $E(4^{+})=1.824$ MeV. The
energy splitting goes in the right direction. The values of the $g$-factors
are $g(2^{+})=1.927$ and $g(4^{+})=1.091$. We see that $g(4^{+})$
is still close to 1 but $g(2^{+})$ becomes very large. This is because
the $g$-factor of a $p_{3/2}$ proton (2.529)is much larger than
that of an $f_{5/2}$ proton (0.3448). Still it might not be unreasonable
to say that the SDI calculation is suggestive of what one should look
for in more realistic calculations.

Talmi cites in his 1993 book \cite{Talmi} another example, the $g_{7/2}$-$d_{5/2}$
space where the single particle splittings are also small. He also
has a nice discussion (p. 445) of pseudo-orbital angular momentum.

\section{Large Space Shell model calculations}

In the work of Kumbatrzki et al.\cite{key-1} large space shell model reuslts
were presentes for $^{86}$Kr energy levels and g factors. The interaction
consisted of monopole corrected G matrix elements based on the CDBonn
interaction of Sieja et al. \cite{key-2}.The results for the g factors
were close to experimental ones. The theory(experimet) values of the
energy levels were 1.613 (1.655) MeV and 2.265 (2.250) MeV for J=2$^{+}$
and 4$^{+}$ respectively.The respecive g factors were 1.03(1.10(5))
and 0.99(1.03 (14))In these calculations only proton valence partices
were active whilst the neutrons formed a closed N=50 core. The protons
were allowed to be in 4 shells 1f$_{5/2}$, 1p$_{3/2}$, 1p$_{1/2}$
and 0g$_{9/2}$. The value of g$_{s}$ was quenched by a factor of
0.75.

We now address and expand upon what was not previously considered in \cite{key-1} ,
how much of the g factor comes from the orbital part of the M1 operator
and how much comes fom the spin. We do this with the results from both the calculations in \cite{key-1}, \cite{key-4} but also examine two additional effective interactions in the same model space, JUN45\cite{jun45} and jj4b\cite{jj4b}.  The results for these interactions and for the wavefunctions are presented in Table \ref{tab:table1}.

\begin{table}
\begin{center}
\caption{g-factors and wavefunction configurations for $2_1^+$ and $4_1^+$ in large scale shell model calculations.}
\label{tab:table1}
\begin{tabular}{| c | c | c | c | c|} \hline
Interaction & blah\cite{key-4} & JUN45 & JJ4B & SDI interaction  \\ \hline
g($2_1^+$) (total,orbital,spin) & 1.03, 0.99, 0.04 & 1.22, 0.93, 0.29 & 1.09 ,0.97,0.12& 1,1,0\\ \hline
g($4_1^+$) (total,orbital,spin) & 0.99, 1.00, -0.01 & 1.03, 0.99,0.04  &1.03 ,0.99,0.04& 1,1,0 \\ \hline
Wavefunction of $2_1^+$ &  &  &   \\ \hline
3410 & 0.05 &<0.05  &   0.07 &\\ \hline
4202 & 0.06 & <0.05 &  0.06 &\\ \hline 
4220 & 0.08 & <0.05 &  <0.05 &\\ \hline
4310 & 0.13 & <0.05 &  0.07 &\\ \hline
4400 & 0.20 & 0.06 &   0.19 & 0.48 \\ \hline
5210 & <0.05 & 0.05 &  <0.05 &\\ \hline
5300 & <0.05 & 0.13 & 0.08  & 0.24\\ \hline
6110 & 0.06 & 0.11 &   0.05&\\ \hline
6200 & 0.23 & 0.45 &  0.27 & 0.28\\ \hline
Wavefunction of $4_1^+$ & & & & \\ \hline
5210 & 0.08 & <0.05 & <0.05 & \\ \hline
5300& 0.70 & 0.69  &  0.63 & 0.80\\ \hline
5102 & <0.05 & 0.06 &  <0.05& \\ \hline
4400 & <0.05 & 0.05  & 0.09   &0.20 \\ \hline
4202 & <0.05 & <0.05 & 0.05 &\\ \hline
\end{tabular}\end{center}
\end{table}

In each case, we find that the g factors are close to 1 and dominated by the orbital contribution.  

In looking at the wavefunctions we see that the wave funcitons are more complicated than those in
the small space of sections 2.

However the combied occupations of orbits not included in the small space , namely p$_{1/2}$and g$_{9/2}$, is not large. We note that for the 4$^{+}$ state b$^{2}$=0.799.
in the small space In the large space the weight ranges from 0.63 to  0.70 , not so
different. 

This suggests that the simple model in the previous section is not
an unreasonable starting point. Of course ultimately large space calculaiotns
have to be performed to compare with the data.

We also have results for the second 2$^{+}$ state where the g factor in the case of the JUN45 interaction continues to be orbital dominated, 1.0596 (0.9873,0.1830). In a vibrational collective
model one can otain such a result--orbital dominance ,where the 2$^{+}$
and 4$^{+}$ states are part of a 2 phonon triplet.  

We thus have 2 types of models which lead to orbital dominance--one
a schematic SDI calculation in a small space and a more realistic
calculation in a larger space. One important lesson in all this is
that one does not have to have pure LS wave functions to get results where
almost all the contributions are orbital.

We are grateful to Igal Talmi for valuable comments and especially
to Kamila Sieja for generously providing us with the large scale shell
model results.

We thank Gerfried Kumbartzki, Noemie Koller and Karl -Heinz Speidel
for their interest.

Brian Kleszyk also thanks the Rutgers Aresty Research Center for undergraduate
research for support during the 2013-2014 academic year.

\end{document}